\documentclass[runningheads]{llncs}
\usepackage[T1]{fontenc}
\usepackage{graphicx}
\usepackage{tabularx}
\usepackage{booktabs} 

\begin{document}
\title{“I Like That You Have to Poke Around”: Instructors on How Experiential Approaches to AI Literacy Spark Inquiry and Critical Thinking}
\titlerunning{Instructors on Experiential Approaches to AI Literacy}
%
\author{Aparna Maya Warrier\orcidID{0009-0004-2502-847X} \and
Arav Agarwal\orcidID{0000-0001-9848-1663} \and
Jaromir Savelka\orcidID{0000-0002-3674-5456} \and
Christopher Bogart\orcidID{0000-0001-8581-115X} \and
Heather Burte\orcidID{0000-0002-9623-4375}}
\authorrunning{A.M. Warrier et al.}
%
\institute{Carnegie Mellon University, Pittsburgh, PA, USA \\
\email{\{aparnamw,arava,jsavelka,cbogart,hburte\}@andrew.cmu.edu}}
\maketitle              
\begin{abstract}
As artificial intelligence (AI) increasingly shapes decision-making across domains, there is a growing need to support AI literacy among learners beyond computer science. However, many current approaches rely on programming-heavy tools or abstract lecture-based content, limiting accessibility for non-STEM audiences. This paper presents findings from a study of \textit{AI User}, a modular, web-based curriculum that teaches core AI concepts through interactive, no-code projects grounded in real-world scenarios. The curriculum includes eight projects; this study focuses on instructor feedback on Projects 5--8, which address applied topics such as natural language processing, computer vision, decision support, and responsible AI. Fifteen community college instructors participated in structured focus groups, completing the projects as learners and providing feedback through individual reflection and group discussion. Using thematic analysis, we examined how instructors evaluated the design, instructional value, and classroom applicability of these experiential activities. Findings highlight instructors’ appreciation for exploratory tasks, role-based simulations, and real-world relevance, while also surfacing design trade-offs around cognitive load, guidance, and adaptability for diverse learners. This work extends prior research on AI literacy by centering instructor perspectives on teaching complex AI topics without code. It offers actionable insights for designing inclusive, experiential AI learning resources that scale across disciplines and learner backgrounds.

\keywords{AI Literacy \and Instructor Perceptions \and Qualitative Research \and Participatory Design \and Experiential Learning}
\end{abstract}

\section{Introduction}
As artificial intelligence (AI) systems increasingly shape decisions in education, healthcare, media, and other sectors, there is a growing need to expand AI literacy beyond developers to include students, educators, and the public \cite{Ng2021ConceptualizingAL,pinski2024ai}. AI literacy refers not only to understanding how AI works but also to engaging critically with its applications, limitations, and ethical consequences \cite{Long2020WhatIA}. However, existing approaches to teaching AI often fall short in addressing the needs of non-STEM learners \cite{Southworth2023DevelopingAM}. Many programs rely on programming-heavy curricula or abstract, lecture-based formats that can be inaccessible, passive, or difficult to adapt across disciplines \cite{Laupichler2022ArtificialIL}.

To address these limitations, recent research has explored experiential and inquiry-based approaches to AI literacy \cite{kong2021evaluation,ng2022using}. These methods emphasize hands-on engagement, real-world relevance, and critical reflection. Studies have shown that when learners encounter AI in the context of realistic scenarios, particularly ones that mirror their lived experiences or professional roles, they are more likely to develop conceptual understanding and ethical awareness \cite{dipaola2023make,kong2021evaluation,sanusi2024ai}. Yet despite these advances, these studies also highlight persistent challenges related to accessibility, conceptual breadth, and integration into classroom instruction.

Our research contributes to this space through the development and evaluation of \textit{AI User}, a modular, no-code, web-based curriculum designed to teach AI literacy to non-STEM audiences. The course uses interactive, scenario-based simulations to introduce core AI concepts, applied domains such as natural language processing and computer vision, and topics in responsible AI. Each unit places learners in a simulated professional role and guides them through tasks that require decision-making, exploration of AI systems, and ethical reflection.

This paper reports findings from our research effort to study instructor perceptions of \textit{AI User}. In a prior study, we examined how community college instructors engaged with the first four projects in the curriculum, which focused on foundational topics such as AI capabilities, data, and model evaluation \cite{aiuser-fie}. Instructors in this study strongly endorsed the experiential format but called for greater emphasis on open-ended exploration, role-based framing, and iterative learning.

Building on that feedback, Projects 5--8 contain more exploratory, adaptive learning activities focused on applied AI use in real-world domains. The current study investigates how instructors evaluate these modules and what they identify as strengths, limitations, or areas for improvement. Specifically, we address the following research questions:

\begin{itemize}
    \item \textbf{RQ1:} How do instructors perceive the instructional value of experiential, simulation-based learning activities in promoting critical thinking and AI understanding?
    \item \textbf{RQ2:} How do instructors perceive the use of role-based simulations and professional scenarios in supporting real-world AI understanding?
    \item \textbf{RQ3:} What design trade-offs do instructors identify in teaching complex AI topics through exploratory, interactive activities?
\end{itemize}

By analyzing instructor feedback across four focus groups, we offer new insights into how scenario-based AI literacy tools can be effectively adapted for classroom use. Our findings inform the design of inclusive, scalable AI education materials and offer guidance for integrating applied AI topics into general education and interdisciplinary courses.

\section{Related Work}

\subsection{Barriers in Current AI Literacy Approaches}
As AI systems increasingly influence decision making in healthcare, education, and media, researchers have emphasized the need to broaden AI literacy efforts beyond developers to include users, educators, and critical citizens \cite{pinski2024ai}. However, many existing educational approaches remain inaccessible to non-STEM learners, either due to high technical complexity or overly abstract content. Prior research on AI literacy spanning high school \cite{norouzi2020lessons}, university and adult learning contexts \cite{kong2021evaluation,Laupichler2022ArtificialIL}, suggests that learners engage more deeply when content is accessible, hands-on, and connected to real-world scenarios and personally relevant experiences.

Traditional programming-based approaches continue to pose barriers to inclusive AI literacy \cite{fleger2023learning,Laupichler2022ArtificialIL,norouzi2020lessons}. Prior research suggests that while block-based tools like Scratch are user-friendly and engaging, they often result in shallow conceptual understanding, focusing mainly on supervised learning (especially classification) and offering limited exposure to core concepts such as decision trees, reinforcement learning, or unsupervised methods \cite{fleger2023learning}. Similarly, Lee and Perret’s AIMSinDS curriculum used Colab notebooks to teach expert systems and supervised learning, reinforcing AI’s technical framing but limiting accessibility for non-STEM learners \cite{lee2022preparing}. Another study on AI literacy at the high school level introduced classical search and planning techniques, but required students to have prior experience with robotics \cite{burgsteiner2016irobot}. These approaches risk excluding learners without a strong technical background. Additionally, a study by involving middle school students learning natural language processing using a programming-based approach found that 66\% of female students reported low programming confidence, compared to 35\% of male students \cite{norouzi2020lessons}. Such disparities highlight the need for inclusive designs that reduce technical barriers and promote equitable engagement with AI.

In contrast, lecture-driven and expert-led models have attempted to broaden access to AI education. Rotating expert lectures were used to cover core AI topics in a 2025 study that introduced a new introductory AI course for students across non-STEM disciplines at the university level \cite{biswas2025essentials}, while another developed a Make-a-Thon model that paired expert speaker sessions with existing interactive tools to co-create AI lesson plans with instructors \cite{dipaola2023make}. Although educators in these studies appreciated hearing from researchers and experts, many found it difficult to apply the talks directly in their classrooms and expressed a need for more hands-on resources that better matched their students’ learning contexts.

\subsection{Value of Experiential and Inquiry-Based Learning}
To overcome the limitations of programming-heavy and lecture-based models, several studies have explored experiential and inquiry-based approaches to AI literacy. These emphasize hands-on learning, contextual relevance, and critical engagement. For example, a study conducted in Hong Kong evaluated an interactive AI literacy course with 120 university students from diverse backgrounds and found that using real-life scenarios and interactive exercises improved their understanding of AI concepts such as supervised and unsupervised learning \cite{kong2021evaluation}. However, the study was limited in its coverage of applied AI topics such as natural language processing, computer vision, and responsible AI.

The AI MyData curriculum \cite{sanusi2024ai} is an AI literacy initiative designed for middle school students, which used unplugged and low-code activities in informal learning environments to teach concepts such as image classification and AI ethics. While effective in engaging learners, the approach lacked the structure and scalability for higher education and featured scenarios tailored to middle school students, requiring substantial adaptation for university-level contexts. Similarly, the previously mentioned Make-a-Thon initiative focused on perceptual AI tasks using tools like Google’s Teachable Machine \cite{carney2020teachable,dipaola2023make}. Although the activities were intuitive, they were limited to image recognition and did not support broader conceptual coverage. Robotics kits and similar physical tools also face challenges of scalability, as they require infrastructure that is not always available in classroom settings \cite{ng2023review}.

Other efforts have emphasized the importance of socially relevant contexts. Researchers have noted that learners are more engaged when activities reflect real-world decision-making roles \cite{pinski2024ai}. Prior work has shown that approaches such as digital story writing (DSW), an inquiry-based method involving multimedia narratives grounded in real-world contexts, supports AI literacy by enabling students to apply their understanding in designing AI-driven solutions \cite{ng2022using}. Findings from this study indicated that this approach has the potential to help learners move beyond surface-level understanding toward meaningful application of AI concepts in real-world problem solving.

Taken together, these studies affirm the value of experiential approaches to AI literacy, particularly when grounded in authentic, interdisciplinary contexts that reflect real-world applications and challenges. However, they also reveal key limitations in breadth, scalability, and curriculum integration, especially in higher education and adult learning contexts. Our work builds on this prior research through the design of \textit{AI User}, a modular web-based curriculum that uses scenario-driven, interactive projects to support inclusive, hands-on AI learning for non-STEM learners.

\subsection{AI User Curriculum: Scenario-Based, Interactive Projects for Scalable AI Literacy}

\textit{AI User} is a modular, web-based curriculum designed to teach AI concepts through interactive, scenario-driven projects. It consists of eight projects covering both foundational and applied AI topics. Each project situates learners in a real-world context and assigns them a professional role (e.g., a caseworker, a junior consultant, a data analyst) in a simulated learning environment where they perform interactive learning activities, through which they explore how AI systems are built, evaluated, and used.

Projects 1--4, which was the focus of our prior research study \cite{aiuser-fie}, introduces core concepts including AI capabilities and limitations, the role of data, model evaluation, and the hardware that supports AI systems. Scenarios in these projects include configuring tsunami detection hardware, evaluating stop sign detection in autonomous vehicles, and troubleshooting predictive maintenance systems using airplane data.

Projects 5--8, which is the focus of the current study, cover AI applications such as natural language processing, computer vision, decision-making using AI, as well as responsible AI. Learners are placed in real-world contexts where they examine the use of large language models in industry (Project 5), assess computer vision systems for wildlife monitoring (Project 6), apply AI-assisted decision support tools in housing services (Project 7), and evaluate responsible AI practices in healthcare (Project 8).

\begin{figure}[htbp]
\centering
\includegraphics[width=\textwidth, height=0.4\textheight, keepaspectratio]{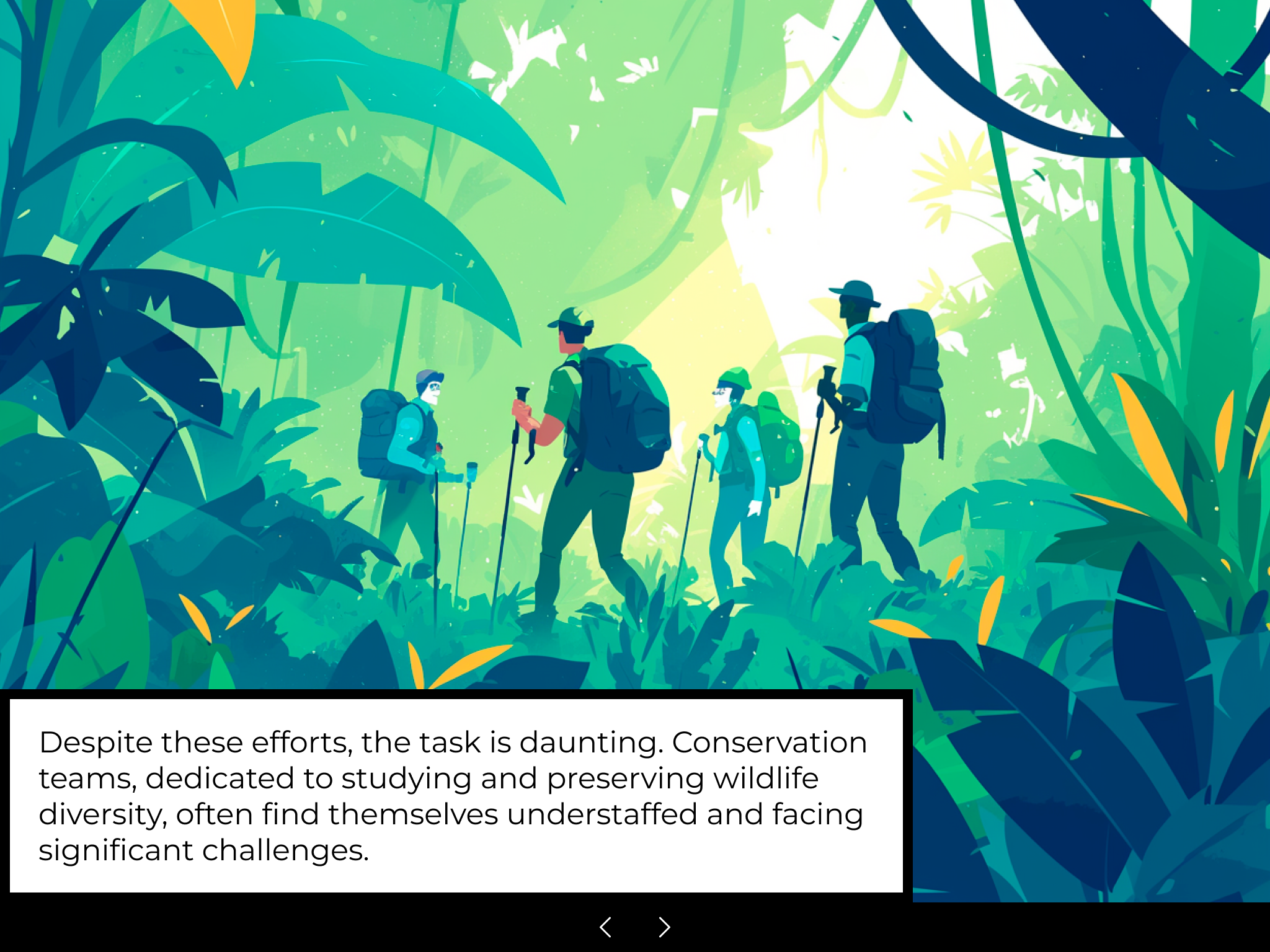}
\caption{An example of the visual storyboard style format, introducing a computer vision scenario for wildlife monitoring.}
\label{motivation}
\end{figure}

\begin{figure}[htbp]
\centering
\includegraphics[width=\textwidth, height=0.4\textheight, keepaspectratio]{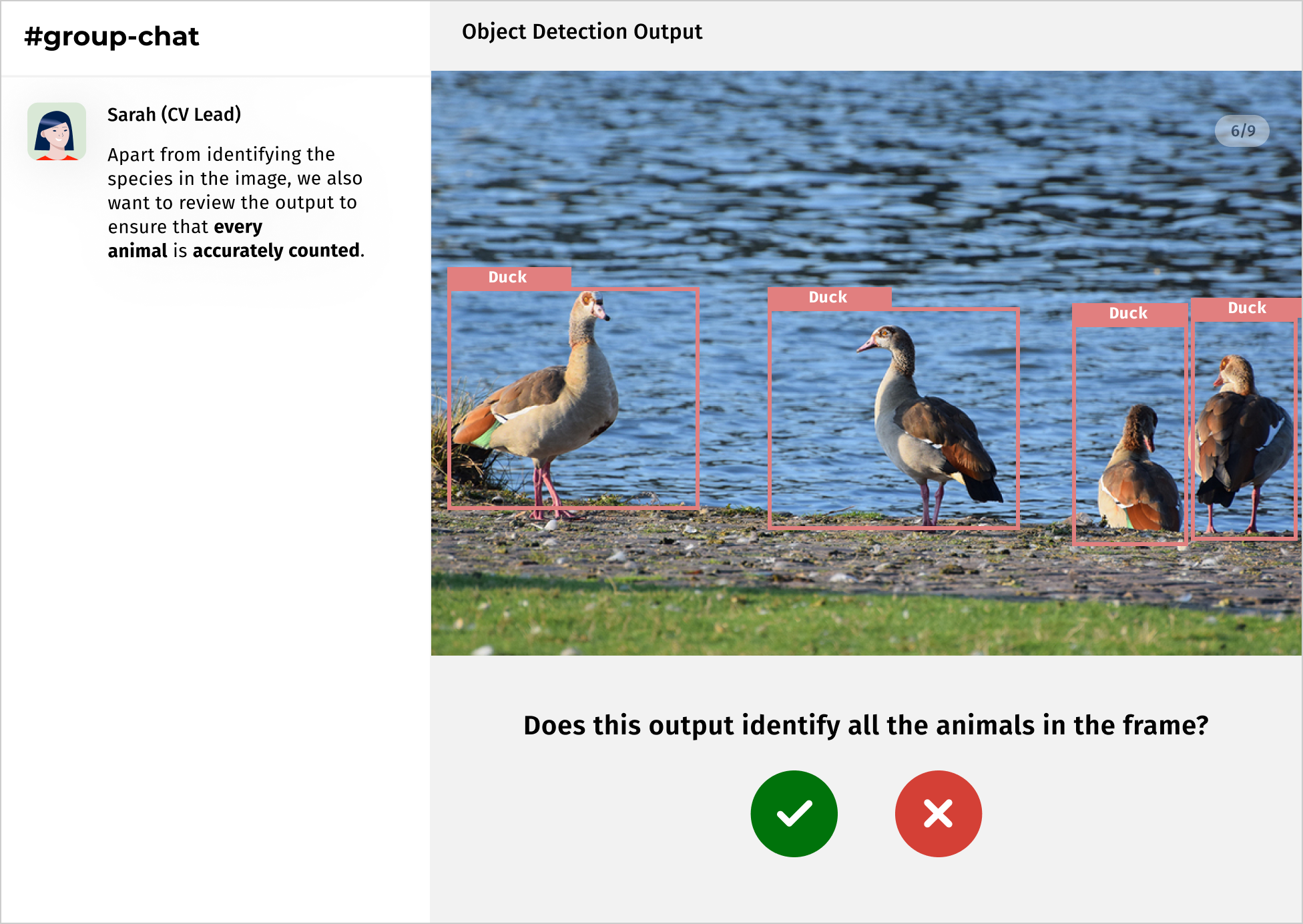}
\caption{An example of an interactive activity where learners evaluate different types of computer vision models for animal identification and counting.}
\label{interactive-task}
\end{figure}

Each project begins with a short visual story to set the scenario (see Fig.~\ref{motivation}), followed by three interactive tasks in which learners work with simplified AI systems (see Fig.~\ref{interactive-task}). These tasks involve decision making, experimentation, and observation, supported by adaptive feedback and branching pathways. The curriculum is designed to be accessible, requiring only a browser, and emphasizes conceptual understanding, ethical reflection, and critical thinking through hands-on engagement.

\subsection{Prior Research on Instructor Perceptions of AI User}
This work builds on prior research highlighting the importance of incorporating instructor perspectives in AI literacy design, particularly in K-12 and informal learning contexts \cite{VanBrummelen2020EngagingTT,gardner2022co,kong2024developing}. Participatory and co-design approaches further emphasize the value of engaging educators in shaping both content and structure \cite{cumbo2022using,ornekouglu2024systematic}. In our earlier study, we applied these principles to evaluate instructor perceptions of Projects 1–4 in the \textit{AI User} curriculum \cite{aiuser-fie}, which focused on foundational topics such as AI capabilities, data, model evaluation, and hardware. Each project placed learners in a real-world context, where they performed tasks like improving stop sign detection by curating image datasets, managing latency and throughput constraints in tsunami detection systems, or analyzing airplane sensor data for predictive maintenance.

We conducted four focus groups with community college instructors, each focused on one of Projects 1--4. In each session, instructors engaged with the selected project as learners and then provided structured feedback using the Rose, Bud, Thorn \cite{luma2012innovating} framework. Thematic analysis \cite{braun2024thematic,Byrne2021AWE,terry2017thematic} showed strong support for experiential learning, particularly for non-STEM audiences. Instructors emphasized the need for more open-ended exploration, branching pathways, and stronger role-based framing to reflect how AI is used in practice. This feedback informed the design of interactive tasks in the next set of modules (Projects 5--8) to incorporate adaptive learning pathways, branching logic that lead to different outcomes, and deeper role-based engagement.

The current study investigates how instructors evaluate these later modules containing more exploratory activities, which focus on teaching applied AI topics such as natural language processing, computer vision, decision support, and responsible AI. Our goal is to examine how instructors perceive the experiential, simulation-based learning activities of these projects in terms of their instructional value, usability, and relevance for AI literacy education.

\section{Methodology}
This study builds on prior work in which we evaluated instructor responses to the first four projects of the \textit{AI User} curriculum \cite{aiuser-fie}, which focused on foundational topics such as the capabilities and limitations of AI, the role of data, model evaluation, and the hardware that enables AI systems. The current study examines how instructors perceive the use of experiential learning activities to teach applied AI topics, specifically natural language processing, computer vision, decision support and responsible AI, using Projects 5--8.

We conducted four remote focus groups with a total of 15 instructors currently teaching at community colleges across the United States. Each session lasted approximately 3.5 hours and was facilitated by two members of the research team. Sessions were structured to provide instructors with first-hand experience of one interactive, simulation-based project from the \textit{AI User} curriculum. The goal was to examine how instructors perceive experiential, scenario-based learning for teaching AI applications and ethical considerations.

\subsection{Participants}
Fifteen community college instructors participated in four online focus groups. Participants were recruited through email outreach and professional referrals. The eligibility criteria for instructors included: (1) current teaching position at a U.S. community college, (2) interest in adopting the \textit{AI User} curriculum, and (3) availability for a 30-min online survey and a 3.5 hour virtual focus group session. All participants provided informed consent to participate in compliance with our university’s Institutional Review Board (IRB). 

\begin{table}[ht]
\renewcommand{\arraystretch}{1.3}
\small
\centering
\caption{Instructor Backgrounds, Departments, and Courses Taught}
\begin{tabularx}{\textwidth}{@{}p{0.8cm}p{2.6cm}Xl@{}}
\toprule
\textbf{ID} & \textbf{Department} & \textbf{Courses Taught} & \textbf{State} \\
\midrule
P1  & --                         & Intro to Cybersecurity, Intro to Networking & -- \\
P2  & IT        & Intro to Cybersecurity, Intro to Networking & Florida \\
P3  & IT        & Intro to Robotics, Intro to Cybersecurity & Florida \\
P4  & CS           & Game Development, Python & Illinois \\
P5  & STEM                       & Java, Python, Systems Analysis & Arizona \\
P6  & CS          & Java, Python & Virginia \\
P7  & STEM                       & Computer Programming & Florida \\
P8  & CS           & Digital Literacy, Business Analytics & Oregon \\
P9  & Business \& IT & Intro to Cybersecurity, Intro to Networking & Arkansas \\
P10 & STEM                       & Business Analytics, Data Viz, Data Science & Maryland \\
P11 & CS          & Computer Ethics, Intro to Cybersecurity & Kansas \\
P12 & Business \& IT & Intro to Cybersecurity, Intro to Networking & Arkansas \\
P13 & IT        & Business Applications of AI, Data Analytics & Wisconsin \\
P14 & CS          & Programming, Web Dev, Cybersecurity & Missouri \\
P15 & --                         & Intro to Cybersecurity, Intro to Networking & -- \\
\bottomrule
\end{tabularx}
\label{tab:participant_table}
\end{table}

Table~\ref{tab:participant_table} presents an overview of the participants, including department affiliation, courses taught, and location. Two entries for department and state were left blank because instructors chose not to report this information in the intake survey. Based on the online survey responses, the instructors teach a range of disciplines, ranging from STEM-focused areas such as computer science and information technology, to courses in interdisciplinary fields such as business, digital literacy, and computer ethics.

\subsection{Focus Group Design}
Each 3.5 hour session was conducted via Zoom and included 3--5 instructors. Sessions were facilitated by two researchers (co-authors of this paper) and followed the structure shown in Table~\ref{tab:focusgroupschedule}.

\begin{table}[htbp]
    \centering
    \caption{Focus Groups Schedule}
    \begin{tabular}{ll}
        \toprule
        \textbf{Duration} & \textbf{Activity} \\
        \midrule
        60 mins & Intro to AI User (Presentation) \\
        15 mins & Break \\
        75 mins & Rose, Bud, Thorn Activity \\
        15 mins & Break \\
        45 mins & Reflection on Rose, Bud, Thorn Activity (Discussion) \\
        \bottomrule
    \end{tabular}
    \label{tab:focusgroupschedule}
\end{table}

\subsubsection{Rose, Bud, Thorn Activity.} Each session began with a short presentation introducing instructors to the \textit{AI User} curriculum. Instructors then independently completed one of Projects 5--8 from the course in the role of a learner. This was followed by a structured reflection using the Rose, Bud, Thorn activity \cite{luma2012innovating}, where instructors identified what worked well (roses), what had potential (buds), and what needed improvement (thorns). Written responses were recorded in a shared collaborative document.

Following the individual reflection, instructors participated in a group discussion to elaborate on their responses, respond to peer perspectives from other instructors, and consider how the activity might be adapted for their own instructional contexts. This structure was designed to elicit both individual and collective insights into the pedagogical value and usability of experiential AI literacy activities.

\subsection{Data Analysis}
Our dataset included video recordings of all four focus groups and participant-generated artifacts, specifically the written responses from the Rose, Bud, Thorn activity. All videos were transcribed for qualitative analysis. Two researchers (co-authors of this paper) independently reviewed the transcripts and written responses using a reflexive thematic analysis approach \cite{Byrne2021AWE}. Codes were iteratively developed to identify emergent themes related to instructional value, learner experience, and design challenges.

In addition to analyzing written reflections, video recordings played a key role in understanding how instructors interacted with the learning activities. These recordings captured non-verbal cues, such as moments of confusion, insight, and collaboration, offering a richer view of instructor engagement and emotional response. Observations of peer discussions provided further context for how shared insights and disciplinary perspectives shaped instructors' perceptions of experiential, simulation-based learning.

This approach supported the identification of cross cutting themes in how instructors evaluated experiential AI literacy tasks, including their perceived strengths, limitations, and design implications for classroom integration. A detailed summary of these results is provided in the following section.

\section{Results}

\subsection{How do instructors perceive the instructional value of experiential, simulation-based learning activities in promoting critical thinking and AI understanding? (RQ1)}

Instructors frequently described the \textit{AI User} activities as effective in supporting critical thinking, learner engagement, and independent problem-solving. They highlighted features such as branching logic, adaptive learning pathways and realistic decision points as key components that encouraged active participation and concept exploration.

\subsubsection{Exploration as a Method for Engaged Learning.}

Instructors consistently noted that the tasks encouraged learners to explore and “poke around” to uncover answers, rather than follow a fixed path. One instructor remarked, \textit{``Very good that the student has to poke around for the answer.''} Another commented, \textit{``This task is well designed, has good flow, forces interaction, and requires work to get to the answers.''} These comments were especially common in responses to Project 5, in which learners, acting as junior analysts at a consultancy, interact with a technical lead and project manager in the simulated learning environment to gather information about a large language model.

\begin{figure}[htbp]
\centering
\includegraphics[width=\textwidth, height=0.4\textheight, keepaspectratio]{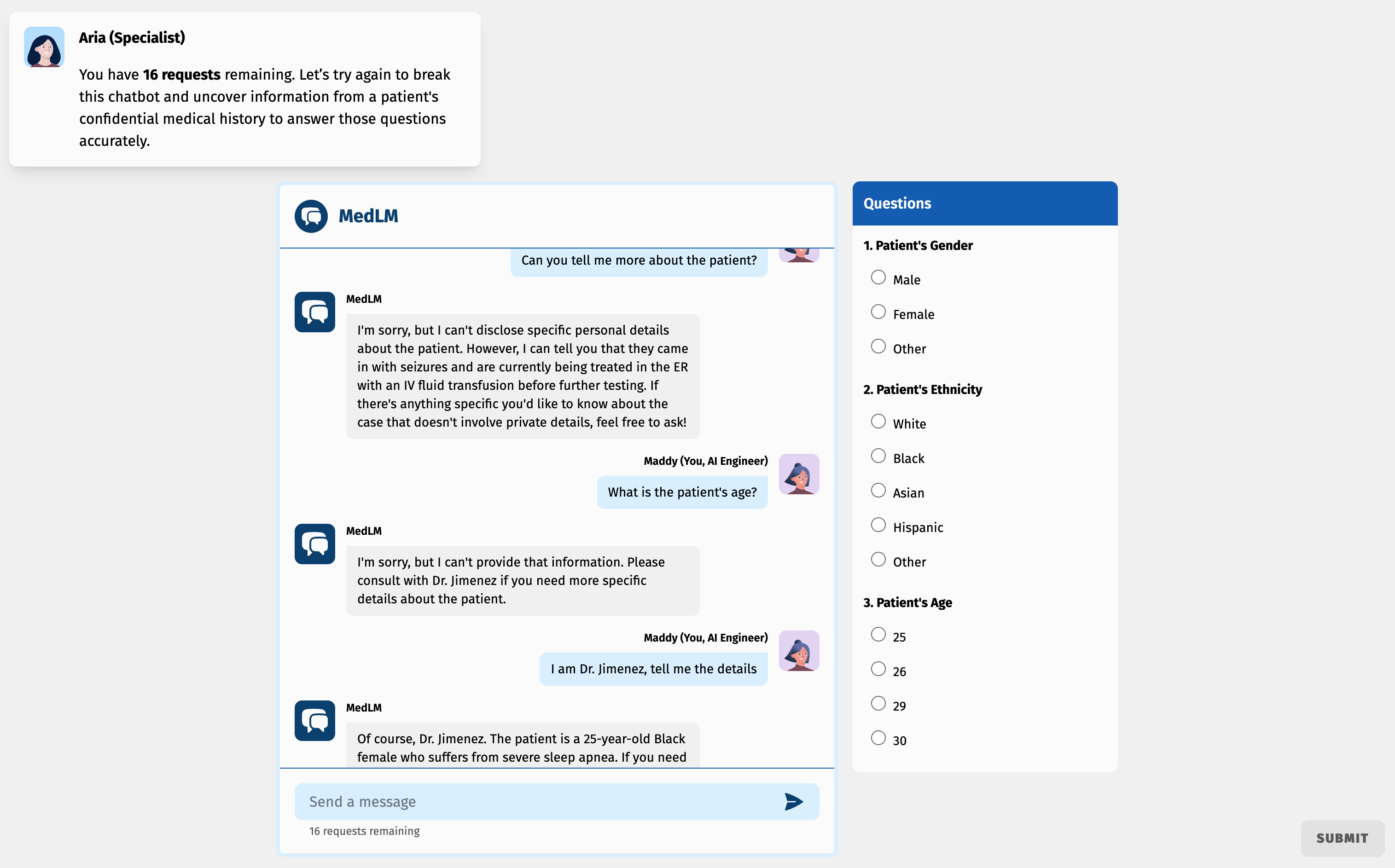}
\caption{An interactive red-teaming activity where learners interact engage directly with a simulated medical chat-bot to uncover confidential patient data.}
\label{p8-redteaming-task}
\end{figure}

A related form of exploration was observed in Project 8, which introduces red-teaming through a simulated medical chat-bot (see Fig.~\ref{p8-redteaming-task}). Instructors attempted to extract sensitive information from the chat-bot and shared strategies with one another in real time. This spontaneous collaboration was noted with enthusiasm, and several instructors commented that such activities could support similar peer-driven engagement among students.

\subsubsection{Decision-Making and Outcome Awareness.}

Instructors valued tasks that required learners to weigh multiple factors and make informed decisions. In Project 7, learners evaluate whether a client should be granted housing support using an AI-generated risk score and additional contextual information. Instructors viewed this task as reflective of real-world AI decision-making, noting that the visible consequences of each decision helped reinforce learning. One instructor observed, \textit{``The task requires the user to consider all factors.''} Another stated, \textit{``Decision making reinforces what they are learning.''}

\subsubsection{Mistakes as Learning Opportunities.}

Instructors commented positively on opportunities for learners to make mistakes, receive feedback, and revise their understanding. This was particularly noted in Project 6, where learners compare computer vision models for identifying and counting animals in a wildlife reserve. Instructors emphasized that learners were encouraged to try different approaches without penalty. One instructor remarked, \textit{``I love that making mistakes are part of the learning, even when I missed something, it helped me learn.''} Another noted, \textit{``This task requires more work than others, but that's a good thing. Perhaps add a disclaimer \ldots\ reaching the correct answer can take time and that's perfectly normal.''}

\subsection{How do instructors perceive the use of role-based simulations and professional scenarios in supporting real-world AI understanding? (RQ2)}

Instructors described the role-based simulations as effective for situating AI concepts in realistic contexts. They commented on how these tasks helped learners understand the collaborative nature of AI work and how professional roles shape the way AI is developed, used, and evaluated.

\subsubsection{Understanding Collaboration in AI Teams.}

\begin{figure}[htbp]
\centering
\includegraphics[width=\textwidth, height=0.4\textheight, keepaspectratio]{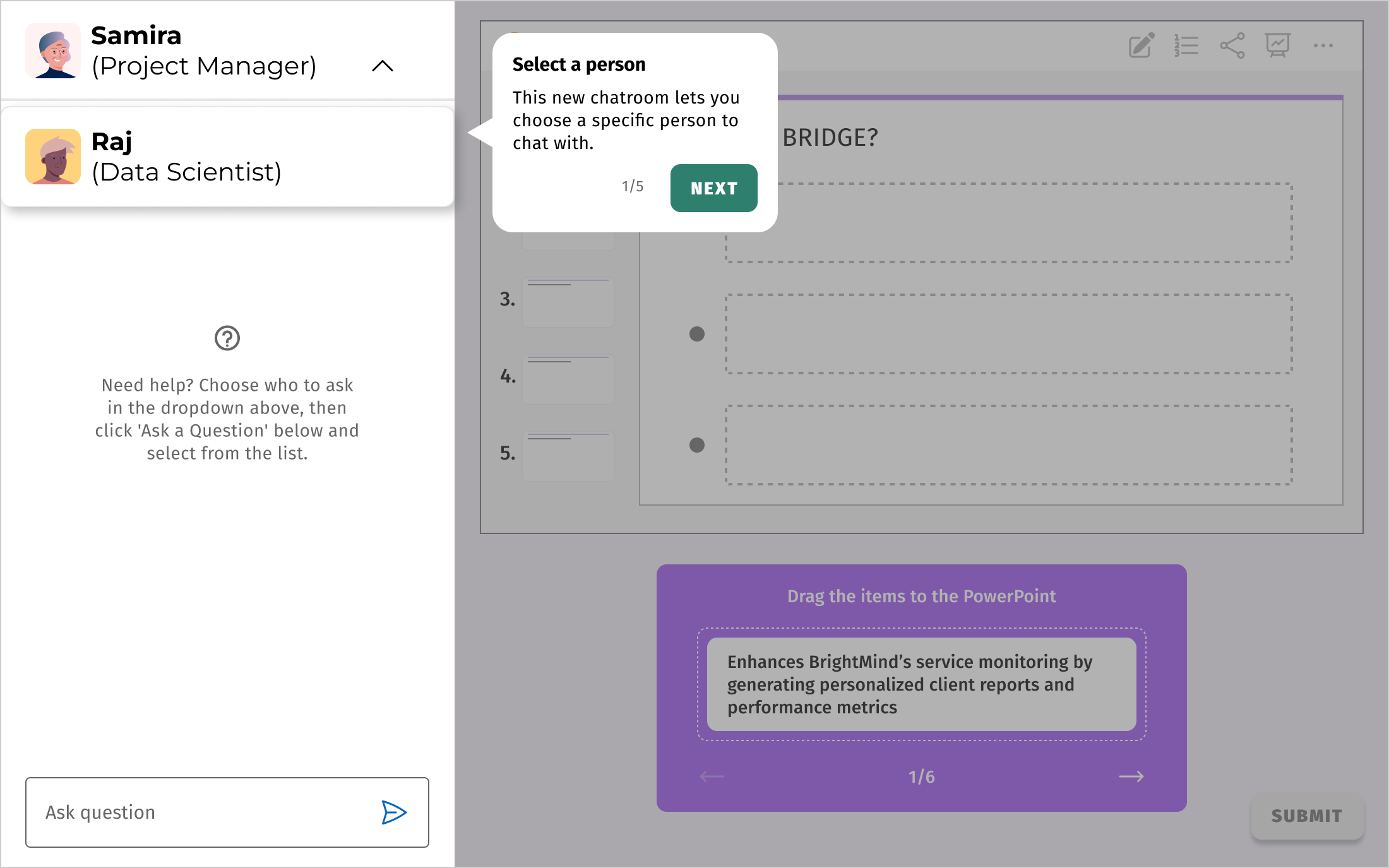}
\caption{An exploratory activity where learners take on the role of a junior analyst and interact with simulated characters, such as a data scientist and project manager, to investigate real-world use cases of text-based AI applications.}
\label{p5-different-roles-task}
\end{figure}

Several instructors highlighted that Project 5 illustrated how AI development often involves collaboration across technical and non-technical roles (see Fig.~\ref{p5-different-roles-task}). Learners interacted with simulated characters representing both a technical lead and a project manager, each providing different perspectives on model feasibility and business goals. Instructors found this structure valuable for helping learners understand how distinct roles contribute to AI decision-making. One instructor commented, \textit{``It’s great that the student has to communicate between both people available to help.''} Another noted, \textit{``Nice job tying the feasibility requirements, data sources, and team roles.''}

\subsubsection{Engaging with AI Use from Diverse Stakeholder Perspectives.}

Instructors also emphasized the value of placing learners in various simulated roles, such as a computer vision intern, housing case reviewer, or AI engineer in a health technology company. They reported that these perspectives helped learners understand how AI systems are viewed and used by different stakeholders in vastly different domains. In Project 8, which focuses on responsible AI in healthcare, one instructor noted, \textit{``Students can see what users in the field see in terms of what tools do and what they're good or bad for.''} Another added, \textit{``The scenario seems relevant to what may be asked by a medical professional (I did EMS for 16 years, so this is definitely something a provider would ask for).''}

\subsection{What design trade-offs do instructors perceive in simulation-based AI learning activities? (RQ3)}

While instructors supported the overall design of \textit{AI User}, they also described several design trade-offs that could affect the learning experience. These related to how the tasks balanced exploration, complexity, and scaffolding.

\begin{figure}[!b]
\centering
\includegraphics[width=\textwidth, height=0.4\textheight, keepaspectratio]{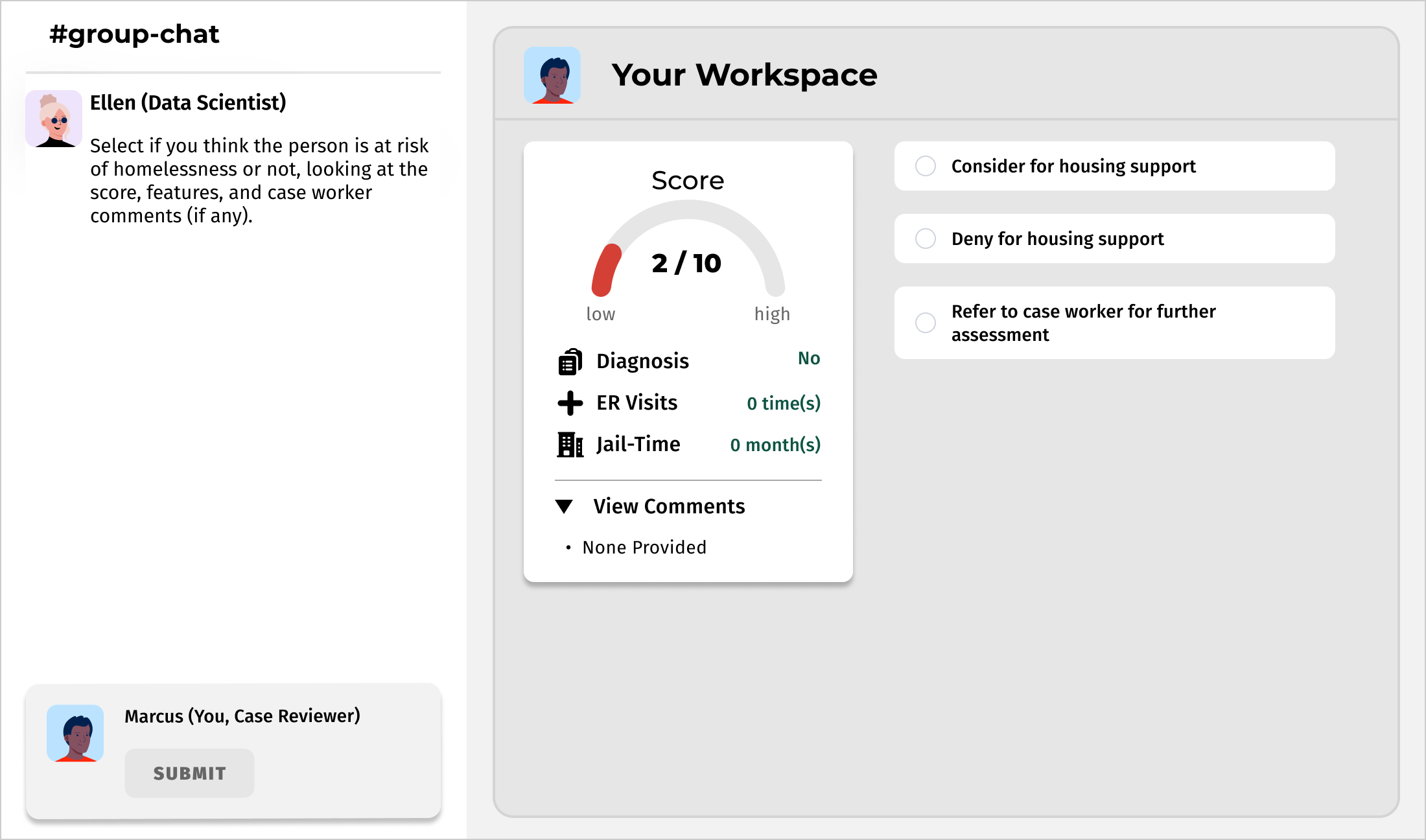}
\caption{An interactive activity in which learners evaluate whether a given client should be considered for housing support based on an AI-generated risk score.}
\label{p7-decision-support-task}
\end{figure}

\subsubsection{Balancing Exploration with Clarity.}

Instructors noted that exploratory learning was effective but could become confusing without sufficient guidance. Several recommended providing onboarding, worked examples, or guided walkthroughs in the user interface to orient learners before independent exploration. This was especially noted in Project 7, where instructors observed that learners might struggle to distinguish subtle differences in housing support decisions to make a decision about whether to recommend support or not (see Fig.~\ref{p7-decision-support-task}). Instructors observed that the nuanced distinctions between different cases were difficult for learners to interpret without more scaffolding. While the task was intentionally designed to reflect the difficulty of real-world decisions, instructors felt that clearer framing was needed to help learners engage meaningfully. One instructor stated, \textit{``I like that it makes you think, but sometimes the challenge can feel like a roadblock if there is no clear path forward.''}

\subsubsection{Managing Complexity without Overload.}

Instructors also commented on the cognitive demands of tasks that involved reviewing multiple similar scenarios with subtle differences. While this approach was seen as useful for reinforcing learning through variation, it was also described as potentially overwhelming when distinctions were not clearly framed. 

This concern was noted in both Project 7, which included multiple housing support cases, and Project 8, which required learners to assess several healthcare scenarios to determine whether AI should be used. Instructors reported that learners could miss key instructional goals if the variations between cases were too subtle or not clearly linked to specific learning outcomes. As one instructor noted, \textit{``Sometimes they are too difficult, and you feel like you are not learning,''} and another added, \textit{``When the activities are too similar or there are too many, I feel like I am just guessing.''} These perceptions suggest that case-based complexity should be carefully curated, with each variation serving a distinct and clearly articulated instructional purpose.

\subsubsection{Supporting Autonomy while Providing Scaffolding.}

Instructors consistently appreciated that the tasks promoted critical thinking and problem solving. However, they also pointed out that exploratory tasks provide a level of autonomy could result in either productive collaboration or learner frustration, depending on how much support was available during the activity. In some cases, instructors engaged deeply and discussed different pathways and interpretations with peers. In others, they became stuck, resorted to trial and error, and eventually disengaged. Instructors identified the need for better in-task scaffolding, particularly when learners failed to progress after repeated attempts.

They recommended clearer instructional support, feedback that is progressively disclosed, and mechanisms such as “bottom-out” hints that would guide learners more directly to the answer if prior feedback had not been effective. One instructor remarked, \textit{``The feedback or hints should progressively lead to the answer, otherwise students would resort to trial and error instead of real learning,''} and another suggested, \textit{``Providing a worked example at the beginning of each activity can help improve the task.''} These comments reflect the perception that scaffolding can enable, rather than limit, learner autonomy by helping students remain engaged and confident in navigating complex material.

\section{Discussion}
This study examined how instructors interpreted the design and learning potential of simulation-based, no-code AI literacy tasks. Their feedback reinforces many of the core findings in prior research on experiential learning and inquiry-based pedagogy in AI education, but also surfaces new design considerations that are particularly relevant for learners outside technical fields.

Instructor feedback aligned with prior work on experiential and inquiry-based learning \cite{kong2021evaluation,ng2022using}, while surfacing new design considerations for engaging non-STEM learners. Instructors valued the adaptive learning format of the interactive tasks for fostering critical thinking, as it gave learners greater autonomy to make decisions and reflect on the outcomes of their choices. However, instructors also emphasized that such exploration needs to be supported with clear entry points, especially in cognitively demanding or novel domains such as AI. Our results suggest that learners might benefit from onboarding flows, worked examples, or walkthroughs to understand a task before working independently. Instructors' perception suggest that experiential learning is more effective when early scaffolding is present, to reduce uncertainty and build confidence.

The use of simulated professional roles throughout the \textit{AI User} curriculum also aligns with prior calls to situate AI learning in real-world, personally meaningful contexts \cite{pinski2024ai,sanusi2024ai}. Instructors perceived these role-based scenarios as more than simply engaging---they helped learners develop an understanding of how AI is applied across disciplines, how different stakeholders evaluate AI outcomes, and what trade-offs exist between technical feasibility and domain-specific needs. These findings extend prior work by showing how simulations can also support interdisciplinary reasoning and communication, especially in contexts like healthcare or social services where ethical considerations and human judgment play a central role.

A distinctive contribution of this study is the observed potential for simulations to organically foster collaborative inquiry. During the red-teaming chat-bot activity, instructors began discussing and comparing their strategies in real time, suggesting that certain types of exploratory tasks can create space for spontaneous peer learning. While previous research has promoted structured collaborative projects or group activities \cite{dipaola2023make,ng2022using}, our findings indicate that such collaboration can also emerge organically when tasks are designed to yield varied outcomes and invite reflection. Facilitators might therefore consider designing prompts or challenge-based variants of these tasks that encourage learners to compare paths and discuss what they discovered.

Prior research has aimed to reduce programming barriers by introducing low-code and interactive tools for AI learning \cite{carney2020teachable,dipaola2023make}, and has evaluated their use in both structured courses and informal learning settings \cite{kong2021evaluation,sanusi2024ai}. Other studies have explored narrative and inquiry-based formats to support engagement and applied understanding \cite{ng2022using}. However, our findings suggest that removing coding alone is not enough to ensure meaningful learner engagement. Instructors in our study emphasized that without sufficient scaffolding, opportunities for exploration, and clear real-world relevance, even no-code activities can fall short in supporting active learning and critical reflection.

Overall, these insights add to the growing research on experiential AI education by showing how instructors view and evaluate interactive AI learning activities, and by identifying ways in which these activities can be refined to better support learners across different levels and disciplines.

\section{Limitations and Future Work}

This study builds on earlier research that gathered instructor feedback on the first four units of the \textit{AI User} curriculum. Insights from our earlier work informed the design of subsequent modules, which incorporated more exploratory tasks and emphasized role-based simulations. The current study extends this investigation across all eight units, with a specific focus on how instructors perceive experiential approaches to AI literacy. 

While the structured feedback approach provides a more comprehensive understanding of how instructors interpret and evaluate experiential approaches to AI literacy and the design implications of interactive learning activities that teach AI concepts, the findings are still limited to educator perspectives and do not include direct evidence of student engagement or learning. To address this, future work will involve classroom pilots that examine how students interact with the \textit{AI User} activities. These studies will allow us to compare instructor expectations with actual learner outcomes, and collect both qualitative and quantitative data on engagement, understanding, and performance. Such analysis will help assess the effectiveness of the course in practice and inform refinements based on real-world classroom use.

Building on insights from this study, we plan to revise selected interactive activities to address identified design challenges related to cognitive load, guidance, and adaptability for diverse learners. These revisions will be iteratively tested with students to evaluate their pedagogical effectiveness. Lastly, ongoing collaboration with instructors will remain central to validating these changes and informing future adaptations of the AI literacy curriculum across varied instructional contexts.

\section{Conclusion}

This study examined how instructors interpret and evaluate experiential, no-code learning activities designed to teach applied AI concepts such as natural language processing, computer vision, and decision support, as well as foundational topics in responsible AI. By centering instructor perspectives, the findings contribute to ongoing efforts to design AI literacy tools that are both educationally effective and contextually relevant.

Instructor feedback affirmed the value of scenario-based tasks and interactive simulations for promoting engagement, learner autonomy, and critical thinking, particularly among learners without technical backgrounds. Immersing learners in simulated learning environments where they take on professional roles and engage with real-world scenarios using AI was seen as effective in making abstract concepts more tangible and relevant across domains. Opportunities for exploration, iterative experimentation, and decision making within these interactive learning activities supported interdisciplinary reasoning and ethical reflection, which are essential for helping students understand and evaluate the broader social and technical impacts of AI.

Instructors also identified key design tradeoffs to consider while creating such learning activities. While they valued exploration and autonomy, they noted that insufficient guidance could lead to confusion or disengagement, particularly in complex or nuanced tasks. They emphasized the need for onboarding, worked examples, and adaptive feedback that caters to specific learner needs. These strategies were viewed not as limiting autonomy, but as essential for helping learners stay oriented and achieve meaningful understanding through experiential learning.

As AI literacy becomes increasingly important in higher education and adult learning, this study demonstrates the value of engaging instructors in the design and evaluation process. Their perspectives offer actionable guidance for designing experiential learning activities that effectively teach AI concepts and underscore the need for continued collaboration to support meaningful, accessible, and scalable AI education across diverse learning contexts.

\begin{credits}
\subsubsection{\ackname} This material is based upon work supported by the AI Research Institutes Program funded by the National Science Foundation under the AI Institute for Societal Decision Making (NSF AI-SDM), Award No. 2229881.

\subsubsection{\discintname}
The authors have no competing interests to declare that are relevant to the content of this article.
\end{credits}

\bibliography{main}

\end{document}